\documentclass[journal]{IEEEtran}
\pdfoutput=1
\usepackage{flushend}
\usepackage{graphicx}
\usepackage{caption}
\usepackage{comment}
\usepackage{xcolor}
\usepackage{amsmath}
\usepackage{amsfonts}
\usepackage{romannum}

\begin{document}
\bstctlcite{IEEEexample:BSTcontrol}
\title{Graph Representation Learning for Wireless Communications}

\author{Maryam~Mohsenivatani, Samad~Ali,
        Vismika~Ranasinghe,
        Nandana~Rajatheva, and~Matti~Latva-Aho

}

\maketitle

\begin{abstract}
Wireless networks are inherently graph-structured, which can be utilized  in graph representation learning to solve complex wireless network optimization problems. In graph representation learning, feature vectors for each entity in the network are calculated such that they capture spatial and temporal dependencies in their local and global neighbourhoods. Graph neural networks (GNNs) are powerful tools to solve these complex problems because of their expressive representation and reasoning power. In this paper, the potential of graph representation learning and GNNs in wireless networks is presented. An overview of graph learning is provided which covers the fundamentals and concepts such as feature design over graphs, GNNs, and their design principles. Potential of graph representation learning in wireless networks is presented via few exemplary use cases and some initial results on the GNN-based access point selection for cell-free massive MIMO systems.
\end{abstract}


\IEEEpeerreviewmaketitle

\section{Introduction}

Wireless networks are shifting from providing connectivity for human-centric end users to internet of connected things and internet of connected intelligence. In order to support novel emerging services driven by societal needs, radical innovations across various layers of communications protocol stack is required. The traditional technologies and methodologies cannot efficiently respond to the emerging large-scale heterogeneity in wireless networks in a real-time fashion as these traditional networking strategies entail complete or quasi-complete awareness of the environment. Hence, novel tools and frameworks are needed to address the shortcomings of the traditional methods that mostly rely on the mathematical realization of the wireless systems. 
Data-driven approaches such as machine learning algorithms have recently shown to be very effective to perform real-time decision making in wireless networks \cite{Ali2020}. However, modern deep learning architectures and solutions such as convolutional neural networks, feedforward networks, and recurrent neural networks are usually well-suited for grid structured and sequential data. On the other hand, many of the datasets in wireless networks have a graph structure and ignoring the inter-relation between the data points leads to the loss of information that could be highly beneficial in training ML algorithms. Therefore, it is highly desirable to use approaches that utilize all the information that exist when modeling the wireless networks as graphs. 

Graphs are ubiquitous non-Euclidean data structures used to characterize complex relationships in all kinds of networks including wireless systems. To enable using deep learning methods on graphs, paradigm of graph representation learning could be used \cite{Hamilton2020}. Graph representation learning is the art of utilizing machine learning methods to extract feature vectors from graph structured data. These feature vectors contain temporal and spatial information of the local and global neighbourhood of each entity in the graph. The representation vector of a node is produced by iteratively aggregating and altering the representation vectors of its neighbors in GNNs. GNNs can be used to only learn the graph feature vector representation or it can be used to solve a downstream prediction task (classification as well as regression) over the given graph in an end-to-end manner. In such an end-to-end learning approach, the feature vectors are learnt as an intermediate step. Moreover, highly expressive powers of GNNs that stems from capturing the full information of the graphs enables them to be extremely effective in solving combinatorial optimization (CO) problems \cite{https://doi.org/10.48550/arxiv.2102.09544}. 

As mentioned earlier, many wireless network datasets are graph structured which could benefit from graph representation learning. This has led to a recent surge of interest in the research community to use GNNs to solve various wireless problems such as topology design, scheduling, link activation and power control to name a few. For example, as an instance of an autoregressive aproach, in \cite{Simsek2021}, authors capitalize on graph embedding and DRL to propose a sequential topology design approach for IAB networks that maximizes the network utility function. To find the optimal topology, IAB-enabled wireless network is modeled as a graph and RL agent state is the current graph (subgraph or partial graph) including activated edges (i.e., links) and the action is the activation of a certain edge that can contribute to the cumulative reward. The effectiveness of the proposed topology formation from various standpoints is then thoroughly presented which highlights the necessity of using the spatio-temporal dependencies through highly expressive graph embeddings. The authors in \cite{Lee2020} put forward a graph embedding based link scheduling framework for device-to-device (D2D) networks eracted as a fully-connected directed graphs to maximize the weighted instantaneous sum rate CO problem without accurate CSI requisite. Likewise, the authors in \cite{Jamshidiha2021} capitalize on a variational graph autoencoder (VGAE) for graph embedding of an interference graph and propose a link activation strategy which is scalable to various network topologies such as D2D networks. A permutation equivariant transferable random edge graph neural networks (REGNNs) followed by a model-free primal-dual learning weight training algorithm for wireless networks is introduced in \cite{Eisen2020}. Shen et al. \cite{Shen2021} address the generic problem of radio resource management problem using message passing GNNs (MPGNNs) without domain-specific knowledge requirement. Distributed machine learning incurs a substantial amount of overhead among heterogeneous agents with heterogeneous data limit. Hence, to settle this issue, a joint Bayesian learning and information-aware graph optimization that enables agents to engage the most informative agents in their own learning process is employed in \cite{Alshammari2020}, which has less communication overhead and faster convergence. Then a unsupervised primal-dual counterfactual optimization, that arises where multiple (often conflicting) tasks and requirements are demanded from the agent, is applied to train the network for decision making that ensures a minimum rate. Two novel GNN based detectors relying upon the expectation propagation (EP) and Bayesian parallel
interference cancellation (BPIC) are proposed in \cite{9832663} to address the problem of uplink multi-user multiple-input multiple-output (MU-MIMO) detection when the multi-user interference (MUI) is strong. 

The representative works show the importance and the power of GNNs in solving complex problems in wireless communication networks. However, there are many other problems in wireless networks that can benefit from graph learning approaches, many of which are not yet studied in the literature. The aim of this paper is to provide an overview of graph representation learning, its potential to solve wireless network problems and provide future research directions for the field. The contribution of this work can be summarised as follows,
\begin{itemize}
    \item Deep learning on graphs as a frontier in deep learning is introduced.
    \item Design of GNNs as an effective framework for graph representation learning is provided with practicalities.
    \item Some applications of GNNs in wireless networks and open challenges and future directions are discussed.
    \item Our findings for GNN-based access point selection for cell-free massive MIMO systems are also presented to embolden the expressive power of GNNs.
\end{itemize}
The organization of the paper is as follows. In Section \Romannum{2} an overview of graph learning is provided. Graph learning applications and relevance to wireless networks is then further discussed in Section \Romannum{3}. Followed by the concluding remarks in Section \Romannum{4}.

\section{Preliminaries on Graphs}\pagenumbering{arabic}
Graphs are a general language for describing and analyzing entities with relations. Graph \(G\) is characterized by a set of nodes \(V\), edges \(E\), and the matrix of initial features of all nodes \(\textbf{X}^{|V|\times d}\) with \(\textbf{x}_v \in \textbf{R}^d\) being the feature vector of a node \(v\) (note that a graph may also have edge features \(\textbf{X}^{e}\) ), i.e. \(G=(V,E,\textbf{X})\). These attached attributes can be weight, ranking, type, sign, or properties depending on the structure of the graph. Let \(v_i\) represent a node of the graph, then the edge from \(v_i\) to \(v_j\) is denoted by \(e_{i,j}=(v_i,v_j) \in E\). The neighborhood of a node \(v\) is defined as \(N(v)=\{u \in V|(v,u) \in E\}\). The adjacency matrix of \(G\) is denoted by  \(\textbf{A}\), which indicates whether pairs of nodes are adjacent or not. For a given undirected graph, \(\textbf{A}\) is a symmetric matrix and for a directed graph symmetry does not hold. \(A_{i,j}=1 \) if  \(e_{i,j} \in E\), otherwise \(A_{i,j}=0 \). If the graph edges have some weights associated to them the adjacency matrix entries can be any real numbers. A graph is extremely sparse if the $|V| \ll |E|$. Edge lists and adjacency lists are also other common ways of representing graphs.

There exist a lot of problems in various industries that require using machine learning methods on graph structured data. Many of these problems could be modeled as node classification, relation prediction, clustering and community detection as well as graph classification. Graph structured data has particular properties that distinguish it from other data types such as images. For instance, in node classification tasks, the nodes in a given graph are not independent and identically distributed (i.i.d.), which is the principal assumption of standard supervised classification. Therefore, there is a need for solutions to handle graph structured data for machine learning tasks. In the next section we provide a brief background on the traditional learning methods over graphs prior to the rise of graph representation learning approaches, which lays the foundation for the modern deep learning methods over graph structured data \cite{Hamilton2020}.

\section{Traditional Learning Methods for Graph Structured Data}
In the traditional methods for handling graph structured data, the graph features or statistics needed for a given task (e.g., node classification) were derived by the domain knowledge or heuristic methods. Then, these extracted features are fed to a standard deep learning model (e.g., a neural network). These feature vectors describe the structural information around a given entity in the graph (local neighbourhood structure). Here, we briefly introduce some of these methods to derive graph statistics (features) at various levels (e.g. node, edge, or graph level) in a given graph \cite{Hamilton2020}.
\paragraph{Node-level Features and statistics}
To obtain nodes features (statistics) of a graph, two general methods can be used : 1) importance-based features such as node degree or various node centrality measures, 2) structure-based features such as node degree, clustering coefficient or graphlet (a small induced subgraph of a large network) count vector. The former category, captures how important a given node is and used to identify the influential nodes in the graph (network). The later, captures structural characteristics of surrounding neighbourhood of a given node and  used to specify the role of a node in a given network.

\paragraph{Edge-level Features and statistics}
Edge feature extraction is basically the feature design for pairs of nodes. In other words, edge-level features quantify the relation between pairs of nodes. Statistical measurements which detect neighborhood overlap are common tools to quantify pairwise relation (similarity). A proximity score between a pair of nodes is calculated. This proximity score can be derived from distance-based methods by only computing the shortest path between two nodes without considering the neighbourhood information. It also can be calculated using the global neighbourhood overlap method that uses the graph structure to find a defined proximity score. Edge-level features are used mostly in link prediction tasks in which the emerging or missing links based on the existing links are predicted.
\paragraph{Graph-level Features and Statistics}
Graph features characterize the entire graph. These features are useful for graph-level classification and comparing two graphs. One way to find graph $G$  feature $\phi(G)$ is by counting different graphlets in the given graph. In other words, given a set of graphlets, we count how many of them exist in a given graph. Instead of extracting explicit feature vectors, implicit kernel methods can be used instead. We do not wish to go into the details of it as it has a rich literature.

Theese methods, in particular node and graph-level features discussed in this section are task dependent and require hand-crafted solutions for building feature vectors. Such a feature extraction approaches are computationally prohibitive as they need to be re-calculated as the task changes. Using such methods is not feasible in problems with large input graphs or problems with combinatorial nature with a very large number of possible outputs. Therefore, an alternative method, graph representation learning is required to tackle such problems. In the following section, graph representation learning (automatically learns the structural features of the graph) as an alternative for traditional learning methods over graphs is introduced in sufficient details.

\section{Graph Representation Learning}
Graph representation learning is automatically extracting task-independent structural features of graphs. In the sequel, we briefly introduce shallow graph representation learning and deep learning on graph data (GNN) for graph representation learning at node level and graph level.
\subsection{Shallow Graph Representation Learning}
Shallow representing learning could be categorized into two groups as follows:
\begin{figure*}[t!]
\centering 
\centering
\captionsetup{justification=centering,margin=2cm}
\includegraphics[width = 18 cm]{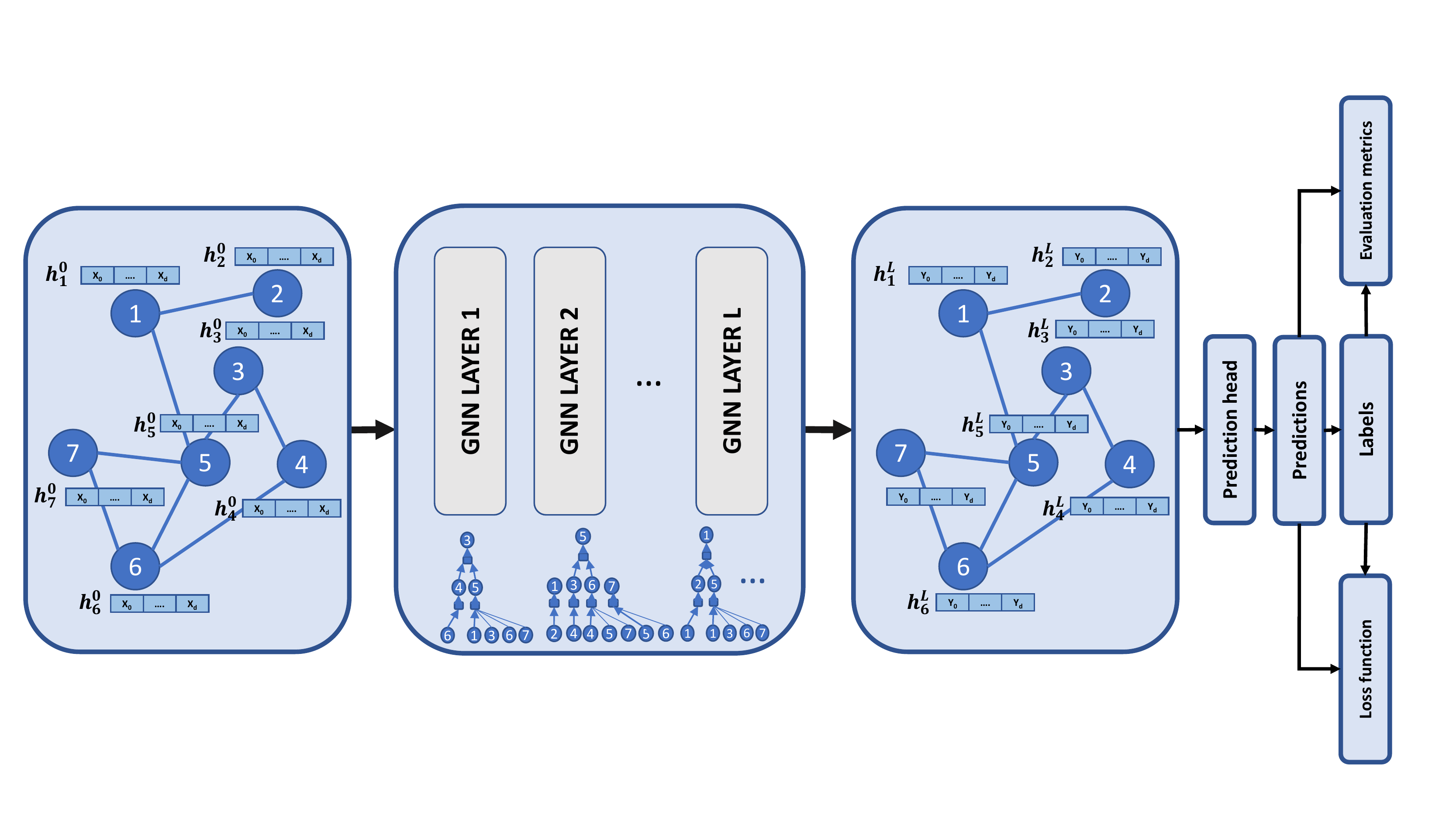}
\caption{GNN training pipeline illustration}
\label{graph_convolution_architecture}
\centering
\end{figure*}

\paragraph{Embedding Nodes} Here we  aim at automatically learning the feature vectors of the nodes such that the properties of the nodes are preserved throughout the whole process of learning (embedding). First approach is Encoder-Decoder approach in which learning node embedding is done with respect to an encoder, a node similarity function, and a decoder design choices. Encoder maps each node to a low-dimensional vector. To do this, a similarity function specifies how the relation in the vector space be mapped to the relations in the original network. Finally, the decoder, maps from embeddings to the similarity score. The goal is to optimize the parameters of the encoder such that the similarity in the embedding space is close to the similarity of the nodes in the original network. Embedding look-up or shallow encoding is the simplest encoding method that directly optimizes an embedding matrix that contains the embedding vectors for all the nodes in the network. Shallow encoder maximizes a similarity score (e.g., the dot product of the feature vectors of two nodes) for a pair of nodes that are similar. DeepWalk and node2vec are two commonly used approaches for shallow encoding which differ in the way that the similarity function is defined. For instance, in DeepWalk approach, the similarity function is the probability that a pair of node co-occur on a random walk on the graph. The objective of random walk is to embed the nodes that appeared on the same random walk together. In a similar fashion, node2vec uses biased random walks to yield the embeddings that can either be biased on the local neighbourhood or the global neighbourhood views in the networks. These methods are unsupervised embedding learning as they do not use any node labels or nodes own features \cite{Hamilton2020}.
\paragraph{Embedding Graphs}
The embedding of the entire graph can be derived by averaging or summation of the embeddings of all the nodes. Graph embedding can also be extracted by making a sub-graph a virtual super-node and finding the embedding of this virtual node and then embedding of all the virtual nodes. Anonymous walk embeddings are similarly a viable candidate to extract the graph embedding.
 
All the above-mentioned algorithms view the graph as a matrix and heavily rely on the matrix algebraic operations. No parameter sharing among the nodes is used to reduce the computational complexity. They are transductive methods and do not use the nodes own features. In the next section, we focus on deep graph representation learning (GNN) that address the shortcoming of the methods that were introduced earlier. 
\subsection{Deep Representation Learning: Graph Neural Networks} 

Conventional deep learning models (e.g., CNNs, RNNs, and feedforward netwoks) cannot be directly used on graph structured data. For example, CNN and RNN are defined for grid and sequence data types, respectively. Since for each node in the given graph a computational graph needs to be used that captures the local neighbourhood structure of that certain node. These computational graphs for all the nodes in the input graph vary in size and topology. One neural network model cannot be applied to all the nodes in a given graph. Therefore, to enable ML learning on graphs, one must design a new deep learning architecture \cite{Hamilton2020}. 

A class of deep learning techniques called graph neural networks (GNNs) has been recently introduced to perform deep learning directly on graph structured data. They can be applied directly to a graph and make predictions and overcome the shortcomings of other approaches discussed in previous sections.

In order to understand how GNNs work, some concepts (design parameters) are defined in the following format. First a computational graph is created for every node that defines the neural structure for that node (it also defines the receptive field of a GNN similar to the receptive field notion in CNNs).
One layer of GNN computes the nodes very own  message or feature on the created computation graph. It also aggregates the messages or features of the neighboring nodes and creates node embeddings. Nonlinearity might also be added to messages or aggregations to increment the expressiveness of the model. A GNN layer compresses multiple vectors into a single vector.
Then multiple GNN layers are stacked sequentially to create a GNN. Graph feature and structure augmentation are performed if the original graph is not used directly as the computational graph to trade off 
computational complexity against accuracy. 
Finally, learning objective  is defined to specify the training of the network (supervised/unsupervised objectives). Different GNNs basically differ in how these operations are done. Classical Graph Convolutional Networks (GCNs) output feature vector for node $v$ at $l$-th layer is calculated by summing the messages of the neighbouring nodes given by
\begin{equation}
    \mathbf{h}_{v}^{(l)}=\sigma\left(\sum_{u \in N(v)} \mathbf{W}^{(l)} \frac{\mathbf{h}_{u}^{(l-1)}}{|N(v)|}\right)
\end{equation}
 where neighbouring node $u$'s message is transformed by $\frac{1}{|N(v)|} \mathbf{W}^{(l)} \mathbf{h}_{u}^{(l-1)}$ where $|N(v)|$ is the cardinality of the neighbouring nodes set, $\mathbf{W}^{(l)}$  is the learnable parameter, and $ \mathbf{h}_{u}^{(l-1)}$ is the messages of the neighbouring nodes. $\sigma$ is the activation function applied to boost the expressiveness of the network.  GraphSAGE is also another common GNN that differs from GCN layer in the fact the GrappgSAGE aggregates node $v$s own message as well by concatenating it with the aggregated messages of the neighbouring nodes. It can be computed as  
 \begin{equation}
     \mathbf{h}_{v}^{(l)}=\sigma\left(\mathbf{W}^{(l)} \left(\mathbf{h}_{v}^{(l-1)} \mathbin\Vert \operatorname{AGG}\left\{\mathbf{h}_{u}^{(l-1)}, \forall u \in N(v)\right\}\right)\right)
 \end{equation}
 $\mathbin\Vert$ represents concatenation operation. $AGG$ can be mean, pool, or long short-term memory (LSTM) functions. To further improve the performance GraphSAGE $l_2$ normalizes the embedding at the end of each layer. Another common and powerful GNN structure is Graph attention networks (GATs). A GAT with $L$ layer returns the embedding 
 \begin{equation}
     \mathbf{h}_{v}^{(l)}=\sigma\left(\sum_{u \in N(v)} \alpha_{v u} \mathbf{W}^{(l)} \mathbf{h}_{u}^{(l-1)}\right)
 \end{equation}
 $ \alpha_{v u}$ is called the attention coefficient reflecting on the importance of the neighboring nodes. Both in GCNs and GraphSAGE all the neighboring node are given the same importance that is not computationally efficient as GNN should allocate more computation power to the more important part of the neighborhood. $ \alpha_{v u}$ is calculated using Attention mechanism. The attention coefficients can be jointly learnt with GNN weight matrices (matrices used for transforming nodes own messages and the one used for neighborhood transformation). GAT is also storage efficient  and has inductive capabilities. $\mathbf{W}^{(l)}$s are learnable parameters that are not regulated by the number of nodes and enable GNN scalability. In other words, as the number of nodes increases $\mathbf{W}^{(l)}$  can be shared and utilized and that is how the problem of different computation graphs with various shapes are handled and the network scales as the new nodes join.

So far we discussed what a GNN layer is. Depending on the objective of the task at hand, GNN layers  are stacked to create a GNN model. Stacking too many layers leads to the over-smoothing problem that means that all the nodes embeddings converge to the same value. Therefore, the embeddings are indistinguishable. This happens due to the fact that the number of layers in GNN indicates receptive field of the nodes neighborhood (in $L$- layer GNN,  every single node has a receptive field of $L$-hop neighborhood). The problem can be tackled by keeping the number of layers small and add to the expressiveness of GNN more and in case large number of layers are required skip connections can be incorporated.

One of the major issues in GNNs that must be handled is that as networks grow large, using original input graph as the computation graph is going to be computation inefficient and inaccurate. The given input graph cannot be the optimal computation graph since it might lack features or it might be too sparse, dense, or large. Hence, feature augmentation or structure augmentation is required depending on the given input graph and the task at hand.

GNN models expressiveness is measured in terms of their capabilities to distinguish different graph structures.
Since GNNs are based on the computational graphs, then the nodes with similar computation graphs will have the same embeddings though they are distinct nodes. To ensure that the GNN is powerful enough to  capture the local neighborhood structure, GNN should use an injective function (every distinct point is mapped to a distinct point) for neighbor aggregation (as it functions over multi-sets) in various layers. Using multi-layer perceptron (MLP) which is an injective multi-set function results in the most expressive GNN among the message-passing GNN category, i.e. Graph Isomorphism Network (GIN).

A GNN training pipeline for node classification is illustrated in Fig. 1. Depending on the task level prediction head harnesses the node embeddings to make predictions. The training is done in either supervised or unsupervised/self-supervised manner. If the labels are given from an external source the training is supervised learning. Otherwise, if the labels/signals are provided by the graph itself, the training is unsupervised. For the classification and regression tasks common loss functions cross entropy and mean squared error (MSE) can be used, respectively. Last but not least, to measure the success of the trained GNN for classification or regression, metrics such as binary classification (accuracy, precision/recall, ROC curves) and root mean square error (RMSE) could be utilized.

To train GNNs, similar to other ML training approaches, the dataset is split into train/validation/test set. The partitioning strategy can span from fixed to random partitioning. As in modern deep learning training pipelines, the training set is used for GNN parameter optimization. Similarly, the validation set is used to to tune the model. The test set is held back and isolated until the training and validation is done. However, the problem with graph-structured data type, as mentioned earlier, is that data leakage happens since the data points are not i.i.d.. The nodes that have been considered for the test phase participate in the node embedding of node that is used for training. To tackle this problem,two different settings could be used. The first approach is that the
whole graph can be used for all splits subject to the partitioning of the labels (not applicable to graph-level tasks).
The second approach is to consider multiple sub-graphs
over the input graph in the dataset
for train/validation/test sets. The later is an inductive solution that can generalize over unobserved scenarios (graphs). The dataset is partitioned in a way to fit the task at hand \cite{Wu2021}.

\section{Graph Representation Learning in Wireless Communications}
Autonomy for robust heterogeneous 6G and beyond wireless networks is conspicuous. Up until now, many machine learning algorithms have been used to solve wireless network problems. However, the majority of ML models (e.g. CNNs, MLPs) do not provide permutation equivariance property which can easily result in intensive training complexity in dense, heterogeneous, and unseen scenarios. This permutation equivariance property of GNNs  guarantees that signal processing with GNNs is independent of labeling of graph signals that implies that relabeling (i.e., applying any permutation and shift) the input signals results in consistent relabelling of the output signal. In other words, the solution to a particular permuted problem is the permuted solution of the original problem. With the rise in agents heterogeneity and magnitude in such networks, having local overview of agents' neighbourhood (i.e., identifying the interactions among agents) becomes essential for each agent in order to contribute to better network planning and management which is delivered by GNNs. Considering the mentioned properties along with their scalability and generalization power, graph learning-based methods spanning from graph filters to GNNs are advantageous to wireless communication networks. In the sequel, we will examine the possible applications of these methods for wireless systems.

\subsection{Cooperative Caching in Mobile Edge Computing(MEC)} With the rise in content-based applications such as video sharing, e-commerce, and social media networking, content caching in networks has already become a key component.This calls for some efficient algorithms that can fulfill low-delay requirement for data fetching. MEC geographic topological structure can be modeled as a graph with the servers as nodes and the relation between them as an edge. Spatio-temporal GNNs are enablers of cooperative caching. Unlike the methods that forecast the popularity of the content and pre-cache the popular content that only consider the temporal features of content popularity, spatio-temporal GNNs also consider the spatial features of content popularity. Another approach could be using graph attention network based self-attention to find representation vectors of the agents in MEC structure. Then these calculated feature vectors are used as the state of a fully decentralized  multi-agent deep reinforcement learning system for cooperative caching \cite{9277917}. The use of GNNs for balancing out between the power consumption and  computation power is also an interesting problem in cooperative caching in MEC yet to be addressed.

\subsection{mmWave Beam Selection}
For transmitter and receiver pair operating at mmWave frequency, the number of antennas at each side is huge so choosing an appropriate pair that fulfills a certain objective becomes an intractable problem. In recent years, ML algorithms proved to be competent enough to address this issue with lower complexity and in a more efficient manner. Considering the fact that the nature of such problems can be further be improved incorporating the spatial topology information into the network, using GNNs is a potential solution to reach an efficient solution within more affordable complexity. For example, \cite{He2022} thoroughly analyzes the problem of joint beam selection and link activation in D2D networks in pursuit of maximizing the total throughput using a GNN-based framework called GBLinks. It is thereby shown that the introduced framework reaches a sub-optimal solution compared to exhaustive search based solutions and also outperforms beam selection strategy with all links active.

\subsection{Packet Duplication}
PDCP packet duplication is an approach that is used to improve the reliability and the latency of the wireless networks. ML methods that are used for packet duplication decisions use various elements such as RSRP and CQI from two or more base stations to the users \cite{9482453}. When such information is fed to a neural network as an input vector, the neural network will not be able to understand the relationship between the different elements of the feature vector. Using graph representation for such data will solve this problem. If the data is formatted such that each base station is one node and connections to the UEs are edges, then the full information about the network architecture could be embedded and used as the input for the neural network that is making PDCP packet duplication decisions.

\subsection{Integrated access and backhaul}
As a result of more available bandwidth in the 5th generation mobile network (5G), a new backhauling technology Integrated Access and Backhaul (IAB), was proposed. IAB enables cell densification and coverage extension by surmounting the need for fibre connections between the small cells and core network which is economically and physically prohibitive. IAB technology uses the same standardized technology for the access and backhaul connection and the resources are also flexibly shared between the access and backhaul over-the-air instead of a prior fixed split of the resources. An IAB network  can be modeled as a graph. One possible case is a graph with IAB nodes or donors as the nodes and the relations between them as edges. IAB networks can have either spanning tree, mesh network, directed acyclic graph(DAG), and etc topologies. topologies. To fully take advantage of the flexibility of IAB networks, efficient topology formation and accordingly dynamic resource allocation are needed as they have major impacts on the system performance. For instance,  the authors in \cite{Simsek2021} investigate a graph-embedding based DRL solution to incorporate the structural information and reduce the complexity of IAB topology formation combinatorial optimization problem. Capitalizing on the aggregated information of a certain node' (either IAB donor or node) neighbours provides insights into latent interactions between that certain node and others as well as indicates the impact intensity of others on a certain node which can bring significant improvements in manifold aspects such as load balancing, anomaly detection, and failure robustness in IAB networks. The above areas are open problems and yet to be explored by the interested researches.

\subsection{Cell-free massive MIMO networks}
Cell-free is a user centric approach where, large number of APs serve a single UE on the same resource block instead of a dedicated AP as in the case of conventional cellular network. Furthermore, these APs a connected to central processing unit (CPU) where the signal processing tasks are performed. It has been shown that cell-free systems outperform small cell networks due to macro diversity achieved by geographically distributed APs serving the same UE. Even though, all the APs  in a system can be used to serve a single user, in practice to achieve scalability best available APs to serve a certain UE must be selected. This selection may depend on the signal strength between UE and APs, front-haul network load which connect APs to the CPU, and interference from other UEs. Since UEs are served non-orthogonally, interference due to use of same pilot sequence for channel estimation, known as pilot contamination, is another key issue which needs to be resolved. GNNs can be used to provide scalable solutions for key problems in cell-free systems like AP selection and Pilot assignment. In \cite{Ranasinghe2021} we have formulated a AP selection problem as two stage classification problem where in the initial stage APs classified based on some metric and then second stage stage classification is formulated as an link prediction task. The main advantage of the proposed GNN based AP selection algorithm is CPU being able to select the potential APs for an UE trying to establish a connection using RSRP measurements of only few APs. Here, as illustrated in Fig. \ref{graph_convolution_architecture}, spatial convolutions based on GraphSAGE inductive framework are performed on two graphs  \cite{Hamilton2020}. In order to make the embeddings independent from features of dynamic UE nodes, UE nodes are connected to APs using only directional edges.  In Fig.  \cite{He2022}, precision and recall curves of the proposed GNN based AP selection algorithm for a cell-free system with 25 APs evenly distributed in a $500 \times 500 \text{m}^2$ are presented.

\begin{figure}[t!]
\includegraphics[width=8cm]{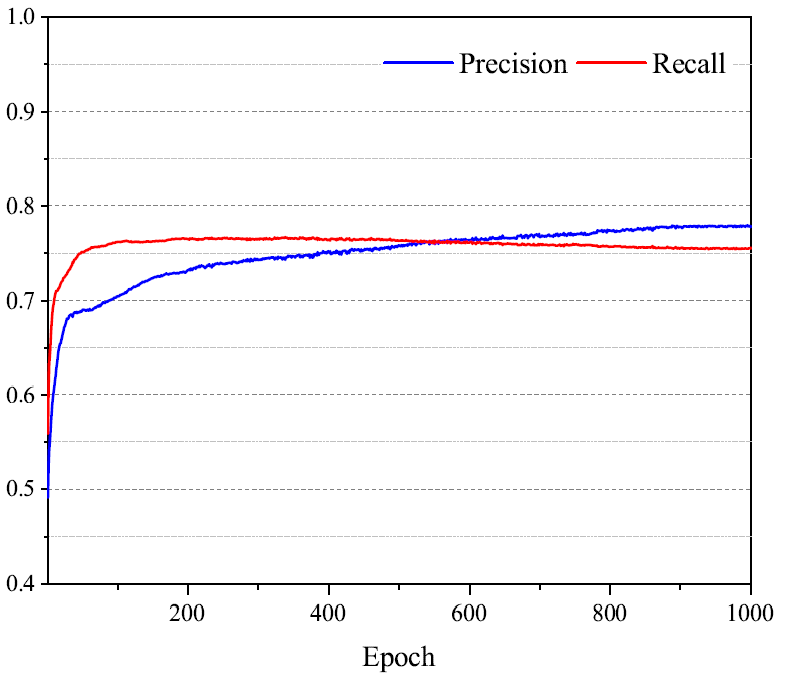}
\caption{Precision and recall of AP selection in a cell-free system with 25 APs}
\label{link_prediction}
\centering
\end{figure}

\section{Conclusion}
In this article, we have studied the potential of incorporating graph representation learning in wireless communication systems. We have reviewed the challenges faced by conventional
ML models for graph-structured wireless networks and discussed the merits
of graph representation learning as an enabling mechanism in tackling them. Then, we have provided an overview of graph representation learning, specifically GNNs, thereafter we presented several  problems in wireless communications that can benefit from these methods. In particular, we have used GNN in access point selection in cell-free massive MIMO and we have analyzed the performance. By way of conclusion, the pivotal role of graph representation learning in wireless communications is undeniable as in other fields. Structural information incorporation and exploitation can significantly contribute to more efficient and accurate predictions. This work can be considered as an essential guide toward solving graph-structured open problems in wireless systems identified earlier for interested researchers.

\bibliographystyle{IEEEtran}
\bibliography{references}
\end{document}